\documentclass[11 pt]{article}
\usepackage{eurosym}
\usepackage{amsfonts}
\usepackage{amsmath}
\usepackage{amssymb}
\usepackage{geometry}
\numberwithin{equation}{section}
\usepackage{cite}
\usepackage{hyperref}

\DeclareRobustCommand{\rchi}{{\mathpalette\irchi\relax}}
\newcommand{\irchi}[2]{\raisebox{\depth}{$#1\chi$}}

\hypersetup{colorlinks=true,linkcolor=blue,urlcolor=blue,citecolor=red}
\input{tcilatex}
\begin{document}

\title{Mean and variance of the cardinality of particles in polyanalytic
Ginibre processes via a  quantization method}
\author{ Zouha\"{i}r Mouayn$^{\ast}$, Mohamed Mahboubi$^{\flat }$ and  Othmane El Moize$^{\flat }$   }
\maketitle
\vspace*{-0.7em}
\begin{center}
\textit{{\footnotesize ${}^{\ast }$ Department of Mathematics, Faculty of Sciences
and Technics (M'Ghila),\vspace*{-0.1em}\\ Sultan Moulay Slimane University, P.O. Box. 523, B\'{e}ni Mellal, Morocco \vspace*{1 em}\\[3pt]
${}^{\flat }$ Department of Mathematics, Faculty of Sciences,\vspace*{-0.2em}\\Ibn Tofa\"{i}l University, P.O. Box. 133, K\'enitra, Morocco
 }
}
\end{center}
\maketitle

\begin{abstract}
We discuss the mean and variance of the number \textquotedblleft
point-particles\textquotedblright\ $\sharp _{D_{R}}$\ inside a disk $D_{R}$
centered at the origin of the complex plane $\mathbb{C}$ and of radius $R>0$ with
respect to a Ginibre-type (polyanalytic) process of index $m\in \mathbb{%
\mathbb{Z}
}_{+}$ by quantizing the phase space $%
\mathbb{C}
$ via a set of generalized coherent states $\left\vert z,m\right\rangle $ of
the harmonic oscillator on $L^{2}\left( 
\mathbb{R}
\right) $. By this procedure, the spectrum of the quantum observable representing  the indicator function $\rchi _{D_{R}}$ of $%
D_{R}$ (viewed as a classical observable) allows to compute the mean value
of $\sharp _{D_{R}}$. The variance of $\sharp _{D_{R}}$ is obtained  as a special eigenvalue of a 
quantum observable involving to the auto-convolution of $\rchi _{D_{R}}.$ By adopting a coherent states
quantization approach, we seek to identify classical observables on $%
\mathbb{C}
,$  whose quantum counterparts may
encode the first cumulants of $\sharp _{D_{R}}$ through spectral properties.
\end{abstract}

\section{Introduction}

\hspace*{1em} Random point processes, which may be defined as distributions over
configurations of points, have arose in many different areas (and still may
emerge in unexpected places), such as statistical mechanics, combinatories,
representation theory and probability theory. They have been used to model
fermions in quantum mechanics, zeros of Gaussian analytic functions, in
classical Ginibre and circular unitary ensembles from random matrix theory,
for examples arising from non-intersecting random walks and random spanning
trees see  \cite{Soch,Hough}. They also have been
used in machine learning \cite{Kulesza-Taskar}, where the state
space is finite and in wireless communication to model the locations of
network nodes \cite{Leo,Miyoshi-Shirai}.\\

Characterizing these processes may be viewed as studying   systems of
interacting \textquotedblleft point-particles\textquotedblright . In this
respect Macchi \cite{Macchi} introduced a time independent model
(called the fermions point process) describing the statistical distribution
of a fermions system in thermal equilibrium, with the property that the $k-$%
point correlation functions have the form of determinants built from a
correlation kernel, implying that particles obey the Pauli exclusion
principle. Such a point process with a kernel function encoding
repulsiveness, exhibit hyperuniformity or rigidity properties \cite{Torquato}, that is the fluctuations of the number of points in a given region is smaller than compared to a Poisson point process with the same
intensity. To predict the large deviation asymptotics for the number of
points in large balls becomes a subject to intense investigating (see\ \cite{Shirai1} as an example) keeping in mind the basic
electrostatic fact stating that the variance of the number of points in a
box should grow like the surface area instead of the volume.\\

The Ginibre point process, in particular, was first introduced in  \cite{Ginibre} and specifically formed by the eigenvalues of a hermitian matrix with standard complex Gaussian entries (also known as the Gaussian Unitary Ensemble GUE). It models the positions of charges of a two-dimensional Coulomb gas in harmonic potential at the temperature $\beta=2$.  At the limit of these eigenvalues appears the infinite analytic  Ginibre process
on $\mathbb{C}$, denoted $\mu _{\exp },$ associated with the reproducing
kernel $e^{z\overline{w}}$ of the Fock-Bargmann space $\mathcal{A}_{0}\left( 
\mathbb{C}\right) $ of entire complex-valued functions which are $%
e^{-\left\vert z\right\vert ^{2}}d\nu -$square integrable on $\mathbb{C}$
, $d\nu$ being the Lebesgue measure on $\mathbb{C=}$ $\mathbb{R}^{2}$. $%
\mu _{\exp }$ arises from the two-dimensional one-component plasma or
jellium model at a special temperature and it is well known to be exactly
solvable \cite{JLM}.\\

By \cite{AIM}, it turns out that the space $\mathcal{A}_{0}\left( 
\mathbb{C}\right) $ coincides with the eigenspace of the magnetic Laplacian $%
\tilde{\Delta}_{1}=-\partial \overline{\partial }+\overline{z}\overline{%
\partial }$ associated with the lowest Landau level $m=0$. The   eigenspace $%
\mathcal{A}_{m}(\mathbb{C}):=\{\varphi \in L^{2}(\mathbb{C},e^{-\left\vert
z\right\vert ^{2}}d\nu );\,\tilde{\Delta}_{1}\varphi =m\varphi \}$
corresponding to  higher Landau levels $m\in \mathbb{Z}_{+}$ also admit
explicit reproducing kernels. For each fixed $m\in \mathbb{Z}_{+}$, the
associated determinantal point process (Ginibre-type) has been considered by
Shirai \cite{Shirai2} who was concerned by the random variable $%
\sharp _{D_{R}}$ counting the number of points inside a disk $D_{R}$
centered at the origin of $\mathbb{C}$ and of radius $R>0$. He proved that
the variance $\mathbb{V}ar(\sharp _{D_{R}})$ can be expressed as an integral
involving the Laguerre polynomial (see Eq.\eqref{eq3.8} below) and
investigated its asymptotic behavior as $R\rightarrow \infty .$ Hyperuniformity of these Ginibre-types processes has been also discussed by Abreu et al \cite{abreu17} in the framework of the so-called infinite Weyl-Heisenberg Ensemble.\\

 In this paper, our purpose is to recover the mean and variance of  $\sharp _{D_{R}}$  from a coherent states quantization view point. We precisely
derive the mean value of \ $\sharp _{D_{R}}$ by summing   eigenvalues of
the quantized quantum counterpart (operator) form of the indicator function $%
\rchi _{D_{R}}$ of the disk $D_{R}$. To obtain the variance we proceed
by quantizing/dequantizing, essentially, the auto-convolution of $\rchi _{D_{R}}.$
The coherent states with respect to which we are performing this
quantization belong to $L^{2}\left( \mathbb{R}\right) $, and are obtained by
displacing \textquotedblleft \textit{\`{a} la Perelomov}\textquotedblright
,\ via the Schr\"{o}dinger representation of the Weyl-Heisenberg group, an
eigenfunction of the harmonic oscillator, given by a Hermite function.  By
regarding this determinantal point process from a coherent states viewpoint
we seek to identify classical observables on the phase space $\mathbb{C},$
attached to the geometry of the disk $D_{R}$, and whose quantum counterpart operators
may encode, through spectral properties, the first cumulants of the random variable $\sharp _{D_{R}}$.\\

 The organization of the paper is as follows. In section 2, we
briefly review the formalism of determinantal point processes. In section 3,
we recall a generalized Bargmann-Fock spaces with their reproducing kernels. We also summarize  the polyanalytic Ginibre process in $\mathbb{C}$ with	 some of its properties. We give a brief review of the standard coherent states in section 4. Section 5 is devoted  to  an explicit quantization of the complex plane via coherent states attached to generalized Bargmann-Fock spaces. In section 6, we compute the mean value and the
variance of the random variable $\sharp_{D_{R}}$. 

\section{Determinantal point processes}

We briefly recall some basic notions on DPP, for more details we refer to \cite{Soch} and references therein.

\smallskip

Let $\mathcal{E}$ be a one-particle space, generally a separable Hausdorff
space, and $B(\mathcal{E})$ the topological Borel $\sigma $-field. We fix a
radon measure $\lambda (dx)$ on $(\mathcal{E},\mathcal{B}(\mathcal{E}))$.
Let $X$ be the space of countable configurations of particles in $\mathcal{E}
$, usually $X$ is called the configuration space. We assume that each
configuration $\xi :=(x_{i})$, $x_{i}\in \mathcal{E}$, $i\in \mathbb{Z}$ or $%
i\in \mathbb{Z}_{+}$ if $d>1$, is locally finite, that is, for every compact
set $K\subset \mathcal{E}$ the number of particles in $K$ is finite. Let $%
B\subset \mathcal{E}$ be a Borel set and define $\mathcal{B}$ as the $\sigma 
$-algebra generated by all cylinder sets $C_{\ell }^{B}=\{\xi \in X:\sharp
_{B}(\xi )=\ell \}$, $\ell \in \mathbb{Z}_{+}$. We call a random point
process (RPP) the triplet $(X,\mathcal{B},P_{r})$ where $P_{r}$ is a
probability measure on $(X,\mathcal{B})$, which may be described by its
correlation functions.\\

Denoting $\mathcal{E}^{k}=\mathcal{E}\times \mathcal{E}\times ...\times 
\mathcal{E}$, a locally integrable function $\rho _{k}:\mathcal{E}%
^{k}\rightarrow \mathbb{R}_{+}$ is called a $k-$point correlation function
for this RPP if for every bounded Borelian disjoint subsets $A_{1},...,A_{n}$
of $\mathcal{E}$ and for every multi-index $(k_{1},...,k_{n})\in (\mathbb{Z}%
_{+})^{n}$ with $k_{1}+...+k_{n}=k$ the following formula holds true 
\begin{equation}
\mathbb{E}\left( \prod_{i=1}^{n}\frac{(\sharp A_{i})!}{(\sharp A_{i}-k_{i})!}%
\right) =\int_{A_{1}^{k_{1}}\times ...\times A_{n}^{k_{n}}}\rho
_{k}(x_{1},...,x_{k})\lambda ^{\otimes }(dx_{1}...dx_{k})  
\end{equation}%
where $\mathbb{E}$ is the mathematical expectation with respect to $P_{r}$.
The problem of existence and uniqueness for a RPP defined by its correlation
functions was studied in \cite{Lenard1, Lenard2}. This is
the case when the distribution of the random variable $\sharp _{A}$ counting
the number of particles inside a bounded Borel set $A$ is uniquely
determined by its moments. \\

A determinantal (or fermions) point process (DPP) with kernel $K(x,y)$ is a
RPP in $\mathcal{E}$ such that its $k$-points correlation function is given
by 
\begin{equation}\label{eq2.2}
\rho _{k}(x_{1},...,x_{k})dx_{1}...dx_{k}=\det (K(x_{i},x_{j}))_{1\leq
i,j\leq k}\lambda ^{\otimes }(dx_{1}...dx_{k}).
\end{equation}

Now let $\mathfrak{K}$ be a self adjoint integral operator on $L^{2}(%
\mathcal{E},\lambda (dx))$ with kernel $K(x,y)$ and suppose that $Spec(%
\mathfrak{K})\subset \lbrack 0,1]$ and $\mathfrak{K}$ is of locally trace
class then there exist a unique DPP associated with kernel $K$ and 
$\lambda (dx)$ such that its  $k$-point correlation function is given by \eqref{eq2.2}. For ${D}\subset \mathcal{E}$ a relatively compact, the random variable $\sharp _{{D}}$, in this DPP, has the
same distribution as a sum of independent Bernoulli random variables $%
(\lambda _{i}^{{D}})_{i}$, where $\lambda _{i}^{{D}}$ is an
eigenvalue of the restriction to ${D}$ of the operator $\mathfrak{K}$%
. The mean value of $\sharp _{{D}}$ is given by%
\begin{equation}\label{eq2.3}
\mathbb{E}\left( \sharp _{{D}}\right) =\int\limits_{\mathcal{D}%
}K(x,x)\lambda (dx)  
\end{equation}%
and its variance 
\begin{equation}\label{eq2.4}
\mathbb{V}ar\left( \sharp _{D}\right) =\int\limits_{\mathcal{D}}\lambda
(dx)\int\limits_{\mathcal{E\setminus D}}\lambda (dy)\left\vert K\left(
x,y\right) \right\vert ^{2}. 
\end{equation}

\section{Ginibre-type processes}

\subsection{Generalized Bargmann-Fock spaces $\mathcal{A}_{m}(\mathbb{C})$}

The motion of a charged particle in a constant uniform magnetic field in $%
\mathbb{R}^{2}$ with a strength proportional to $B >0$, is described (in
suitable units) by the Schr\"{o}dinger operator 
\begin{equation}\label{eq3.1}
H_{B}:=-\frac{1}{4}\left( (\partial _{x}+iB y)^{2}+(\partial
_{y}-iB x)^{2}\right) -\frac{1}{2}  
\end{equation}%
on $L^{2}(\mathbb{R}^{2},d\nu ).$ By identifying $\mathbb{R}^{2}$ with $%
\mathbb{C}$ in the usual way and setting $d\lambda \left( z\right)
=e^{-|z|^{2}}d\nu \left( z\right) $ then, with the help of the unitary map $%
Q:L^{2}(\mathbb{C},d\nu )\longrightarrow L^{2}(\mathbb{C},d\lambda ),$
defined by $Q[\phi ](z):=e^{\frac{1}{2}B |z|^{2}}\phi (z)$, one can
intertwine the operator \eqref{eq3.1} as $e^{\frac{1}{2}B
|z|^{2}}H_{B }e^{-\frac{1}{2}B |z|^{2}}=\tilde{\Delta}_{B}$. We
take $B =1$ and consider the operator, called magnetic Laplacian 
\begin{equation}
\tilde{\Delta}_{1}:=-\frac{\partial ^{2}}{\partial z\partial \bar{z}}+\bar{z}%
\frac{\partial }{\partial \bar{z}}
\end{equation}%
on $L^{2}(\mathbb{C},d\lambda )$. The spectrum of $\tilde{\Delta}_{1}$
consists of eigenvalues (Euclidean Landau levels) of infinite multiplicity
of the form $\,m\in \mathbb{Z}_{+}.$ In \cite{AIM} generalized
Bargmann-Fock spaces have been introduced as eigenspaces of $\tilde{\Delta}%
_{1}$ as 
\begin{equation}\label{eq3.3}
\mathcal{A}_{m}(\mathbb{C}):=\{\varphi \in L^{2}(\mathbb{C},d\lambda );\,%
\tilde{\Delta}_{1}\varphi =m\varphi \}.  
\end{equation}%
Actually, for $m=0$, $\mathcal{A}_{0}(\mathbb{C})$ turns out to be the
realization by harmonic functions with respect to $\tilde{\Delta}_{1}$ of
the Bargmann-Fock space whose elements are entire functions in $L^{2}(%
\mathbb{C},d\lambda )$. For $m\geq 0,$ $\mathcal{A}_{m}(\mathbb{C})$ turns
out to be the space of true-$m$-polyanalytic functions that is the
orthogonal difference $\mathfrak{F}_{m}(\mathbb{C})\ominus \mathfrak{F}%
_{m-1}(\mathbb{C})$ between two consecutive $m$-polyanalytic spaces
\begin{equation}
\mathfrak{F}_{m}(\mathbb{C}):=\left\{ g\in L^{2}(\mathbb{C},d\lambda ),\ \ 
\overline{\partial }^{m}g=0\right\} .  
\end{equation}%
The eigenspaces \eqref{eq3.3} are pairwise orthogonal in $L^{2}(%
\mathbb{C},d\lambda )$ which decomposes as $\bigoplus_{m=0}^{+\infty }\mathcal{%
A}_{m}(\mathbb{C})$. Furthermore, the reproducing kernel of the eigenspace \eqref{eq3.3} reads $\ $%
\begin{equation}\label{eq3.5}
\widetilde{K}_{m}(z,w)=\pi^{-1}e^{z\bar{w}}L_{m}^{(0)}(|z-w|^{2}) 
\end{equation}%
which may be transfered back into the Hilbert space $L^{2}(\mathbb{C},d\nu )$
in which the kernel \eqref{eq3.5} becomes
\begin{equation}\label{eq3.6}
K_{m}(z,w)=\pi^{-1}e^{-\frac{1}{2}z\overline{z}}\widetilde{K}_{m}(z,w)e^{-\frac{1}{2}%
w\overline{w}}=\left( e^{z\bar{w}-\frac{1}{2}z\overline{z}-\frac{1}{%
2}w\overline{w}}\right) L_{m}^{(0)}(|z-w|^{2}),\text{ }z,w\in \mathbb{C}
\end{equation}%
where $L_{m}^{(\alpha )}(\cdot)$ is the Laguerre polynomial (\cite{Mag}, p.239).

\subsection{The DPP $\protect\mu _{K_{m},d\nu }$}

According to \cite{Shirai2}, we recall the following facts. Let $%
m\in \mathbb{Z}_{+}$ and $\mathfrak{K}_{m}$ be the projection operator
defined on $L^{2}(\mathbb{C},d\nu )$, whose integral kernel is given in \eqref{eq3.6}. The DPP\ associated with $\left( \mathfrak{K}_{m},d\nu
\right) $ is the Ginibre-type point process on $\mathbb{C}$ with index $m,$
denoted by $\mu _{K_{m},d\nu },$ which may also be called polyanalytic
Ginibre point process. Note that $\mu _{K_{m},d\nu }$ is the same as $\mu _{%
\widetilde{K}_{m},d\lambda }$ and is translation and rotation invariant.
According to \eqref{eq2.3}, the first intensity function $\rho
_{1}(z)=K_{m}\left( z,z\right) =\pi ^{-1}$ leads to the expected number of
particles inside the disk $D_{R}$ $=\{\xi \in \mathbb{C},\,|\xi |<R\}$ as%
\begin{equation}
\mathbb{E}_{\mu _{K_{m},d\nu }}\left( \sharp_{D_{R}}\right)
=\int\limits_{D_{R}}\pi ^{-1}d\nu \left( z\right) =R^{2}. 
\end{equation}%
While, by \eqref{eq2.4}, the variance reads 
\begin{equation}\label{eq3.8}
\mathbb{V}ar_{\mu _{K_{m},d\nu }}(\sharp_{D_{R}})=\frac{R}{\pi }%
\int\limits_{0}^{+\infty }e^{-t}\left( L_{m}^{\left( 0\right) }(t)\right)
^{2}\left( \int_{0}^{t\wedge 4R^{2}}\left( 1-\frac{x}{4R^{2}}\right)
^{1/2}x^{-1/2}dx\right) dt
\end{equation}%
which behaves as $C_{m}R$ as $R\rightarrow +\infty $. The constant 
\begin{equation}
C_{m}=\frac{2}{\pi m!}\Gamma \left( m+\frac{3}{2}\right) ._{3}F
_{2}\left( -\frac{1}{2},-\frac{1}{2},m;1,-\frac{1}{2}-m;1\right) \sim \frac{8%
}{\pi ^{2}}\sqrt{m}  
\end{equation}%
as $m\rightarrow +\infty $ (at very higher Landau levels).\\

In \cite{DL}, Demni and Lazag added a geometrical view point to $%
\left( 4.2\right) $ by rewriting it as 
\begin{equation}
Var_{\mu _{K_{m},d\nu }}(\xi (D_{R}))=\frac{1}{\pi ^{2}}\int\limits_{\mathbb{C%
}}e^{-|z|^{2}}\left( L_{m}^{\left( 0\right) }(|z|^{2})\right) ^{2}\left[
Area\left( D_{R}^{c}\cap D_{R}\left( z\right) \right) \right] d\nu (z) 
\tag{3.10}
\end{equation}%
where $D_{R}\left( z\right) $ is the disk centered at $z$ and of radius $R$
and $D_{R}^{c}$ denotes the complementary in $\mathbb{C}$ of $D_{R}.$ If $%
\left\vert z\right\vert $ $\geq 2R$, then $Area\left( D_{R}^{c}\cap
D_{R}\left( z\right) \right) =\pi R^{2}$ since $D_{R}\left( z\right) \subset
D_{R}^{c}.$ Otherwise, if $\left\vert z\right\vert <2R$, then $D_{R}^{c}\cap
D_{R}\left( z\right) $ is the complementary in $D_{R}\left( z\right) $ of
the overlapping of the disks $D_{R}$ and $D_{R}\left( z\right) .$\medskip\\
\textbf{Remark 3.1}. For $\left\vert z\right\vert \leq 2R,$ we may
also write $\ \left[ Area\left( D_{R}^{c}\cap D_{R}\left( z\right) \right) %
\right] /\pi R^{2}=1-\alpha _{R}\left( \left\vert z\right\vert \right) $
where%
\begin{equation}
\alpha _{R}\left( r\right) :=\frac{4}{\pi }\int\limits_{0}^{\arccos \left( 
\frac{1}{2R}r\right) }\sin ^{2}\theta d\theta ,\text{ \ }r=\left\vert
z\right\vert ,
\end{equation}%
is the well known \textit{scaled intersection area} for a circular window of
diameter $2R.$ It is the area of the intersection of two disks each with
radius $R$ and with distance $\left\vert z\right\vert $ between centers$.$
It is known that for $R=2\sqrt{\pi }$, the Fourier transform $\widehat{%
\alpha }_{2\sqrt{\pi }}$ of the radial function $\alpha _{2\sqrt{\pi }}$
generates a translationally invariant DPP, with the kernel given in terms of
the Bessel function $J_{1}$, called a \textit{Fermi-circle }point process in $\mathbb{R}^{2}.$ For more details, see \cite{Torquato}  where the authors analytically provided
an $\mathbb{R}^{n}$ generalization for the unique point process on $\mathbb{R%
}$ to which one can map certain properties of random matrices, fermionic
gases and zeros of the Riemann zeta function.

\section{Coherent states}

Coherent states (CS) were first introduced by E. Schr\"{o}dinger \cite{schro} in order to obtain quantum states in $L^{2}(\mathbb{R})$ that follow the classical flow associated to the harmonic oscillator Hamiltonian 
\begin{equation}\label{HO}
\hat{H}=-\frac{\hslash }{2}\frac{d^{2}}{dx^{2}}+\frac{1}{2}x^{2}-\frac{1}{2}.
\end{equation}
Namely, they are states $\left\vert z\right\rangle \in L^{2}(\mathbb{R})$,
labeled by elements of $\,z\in $ $\mathbb{C}\simeq T^{\ast }\mathbb{R}$ (the
phase space of a particle moving on $\mathbb{R}$) given by 
\begin{equation}
\langle x\left\vert z\right\rangle =\left( e^{z\overline{z}}\right) ^{-\frac{%
1}{2}}\frac{1}{(\pi \hslash )^{4}}\left( -\frac{1}{2\hslash }(\overline{z}%
^{2}+x^{2}-2\sqrt{2}\overline{z}x)\right) ,\text{ \ \ }x\in \mathbb{R}, 
\end{equation}%
$\hslash $ is the Planck parameter (take $\hslash =1$). Their most important
property is the resolution of the identity operator 
\begin{equation}\label{eq4.3}
\mathbf{1}_{L^{2}(\mathbb{R})}=\frac{1}{\pi }\int\limits_{\mathbb{C}}d\nu
(z)\left\vert z\right\rangle \left\langle z\right\vert .  
\end{equation}

The property \eqref{eq4.3} bridges between classical and quantum
mechanics in the sense that every operator acting on $L^{2}(\mathbb{R})$ or
any vector lying there may be decomposed over the phase space $\mathbb{C}$.
The Klauder-Berezin-Toeplitz (or "anti-Wick") quantization, here named
coherent states quantization, of the complex plane consists in associating
to a function $z\mapsto F\left( z,\overline{z}\right) $ (with specific
properties) the operator $P_{F}$ in $L^{2}(\mathbb{R})$ defined by 
\begin{equation}\label{eq4.4}
F\mapsto P_{F}:=\frac{1}{\pi }\int\limits_{\mathbb{C}}d\nu (z)F\left( z,%
\overline{z}\right) \left\vert z\right\rangle \left\langle z\right\vert . 
\end{equation}%
By the operator-valued integral \eqref{eq4.4} we mean the
sesquilinear form%
\begin{equation}
\mathcal{S}_{F}\left( \phi _{1},\phi _{2}\right) =\frac{1}{\pi }\int\limits_{%
\mathbb{C}}d\nu (z)F\left( z,\overline{z}\right) \langle \phi _{1}\left\vert
z\right\rangle \left\langle z\right\vert \phi _{2}\rangle  
\end{equation}%
The form $\mathcal{S}_{F}$ is assumed to be defined on a dense subspace of $%
L^{2}(\mathbb{R})$. \ If $F$ is real-valued and at least semi-bounded, the
Friedrich's extension (\cite{RS}, Vol.II,Th.X.23) $\mathcal{S}_{F}$
allows to define $P_{F}$ as a self-adjoint operator. The function is the
upper (or contravariant) symbol of $P_{F}$, and the mean value of
the latter in the state $\left\vert z\right\rangle $,%
\begin{equation}\label{eq4.6}
z\mapsto \langle z\left\vert P_{F}\right\vert z\rangle =\frac{1}{\pi}\int\limits_{\mathbb{%
C}}d\nu (w)F\left( w,\overline{w}\right) \left\vert \langle z\left\vert
w\right\rangle \right\vert ^{2}=\mathcal{B}\left[ F\right] \left( z\right)
\end{equation}%
is the lower (or covariant) symbol of the operator $P_{F }.$  The map $%
F\mapsto $ $\mathcal{B}\left[ F\right] $ generalizes the Berezin transform.
Following \cite{GO}, a possible criteria for a function $D\ni
z\mapsto $ $F\left( z,\overline{z}\right) $ defined on a certain
domain $D\subseteq \mathbb{C}$ to be considered as a \textquotedblleft
quantizable classical observable\textquotedblright\ via the map \eqref{eq4.6} is that $z\mapsto \mathcal{B}\left[ F\right] \left( z\right) $
be a smooth $\left( \text{i.e. }C^{\infty }\text{ on }D\right) $ with
respect to $\left( q,p\right) \equiv \frac{1}{2}\left( q+ip\right) =z.$ For
more details on the CS quantization, we refer to \cite{Ga, K2, K3, Ber}.

\section{Quantization via CS $\left\vert z,m\right\rangle $}

The Heisenberg group $\mathbb{H}_{1}$ (of degree $1$) is the Lie group whose
underlying manifold is $\mathbb{C\times R=R}^{3}$ with coordinates ($x,y,t)$%
\ and whose group law is $(x,y,t).(x^{\prime },y^{\prime },s)=(x+x^{\prime
},y+y^{\prime },t+s+\frac{1}{2}(xy^{^{\prime }}-x^{\prime }y)).$ The
continuous unitary irreducible representations (UIR) of $\mathbb{H}_{1}$ are
well known (\cite{Fol}, p.37).

\smallskip

Here, we will be concerned with the UIR\ of $\mathbb{H}_{1}$ on the Hilbert
space $L^{2}(\mathbb{R})$ defined by shift and multiplication operators (\cite{Taylor}, \S 1.1) as : 
\begin{equation}
T_{\tau}(x,y,t)\left[ \psi \right] (\xi ):=\exp i\left( \tau t-\sqrt{%
\tau }y\xi +\frac{\tau }{2}xy\right) \psi \left( \xi -\sqrt{\tau }%
x\right) ,\xi \in \mathbb{R}  
\end{equation}%
for $(x,y,t)\in \mathbb{H}_{1}\mathbf{,}$ $\tau >0$ and $\psi \in L^{2}(%
\mathbb{R})$, called the Schr\"{o}dinger representation. This representation  is square
integrable modulo the center $\mathbb{R}$ of $\mathbb{H}_{1}$ and the Borel
section $\sigma _{0}$ of $\mathbb{H}_{1}$ over $\mathbb{R\equiv H}%
_{1}\diagup \mathbb{R}$, which is given by $\sigma _{0}\left( x,y\right)
=\left( x,y,0\right).$ Further, by \cite{BJR} there exists a
self-adjoint, positive semi-invariant operator $\delta $ in $L^{2}\left( 
\mathbb{R}\right) $ such that 
\begin{equation}\label{eq5.2}
\int\limits_{\mathbb{R}^{2}}\left\langle \psi _{1},T_{\tau }(\sigma
_{0}\left( x,y\right) )\left[ \phi _{1}\right] \right\rangle \left\langle
T_{\tau}(\sigma _{0}\left( x,y\right) )\left[ \phi _{2}\right] ,\psi
_{2}\right\rangle d\mu (x,y)=\left\langle \psi _{1},\psi _{2}\right\rangle
\left\langle \delta ^{\frac{1}{2}}\phi _{1},\delta ^{\frac{1}{2}}\phi
_{2}\right\rangle  
\end{equation}
for all $\psi _{1},\psi _{2}\in L^{2}(\mathbb{R},d\xi )$\textit{\ and }$\phi
_{1},\phi _{2}\in Dom\left( \delta ^{\frac{1}{2}}\right).$ The group $%
\mathbb{H}_{1}$ here being unimodular, therefore $\delta $ must be the
identity operator \cite{Duflo}. The connection of the Schr\"{o}dinger representation $T_{\tau}$ with spin $\frac{1}{2}$ representations and magnetic field can be found in \cite{BPS}.\\

Now, according to \cite{Mou3}, CS are obtained
\textquotedblleft \textit{\`{a} la} \textit{Perelomov}\textquotedblright \cite{Perelomov} as orbits of the unitary operator $T_{\tau}$ acting on 
the eigenfunction $\phi _{m}\in $ $L^{2}(\mathbb{R)}$ of the harmonic
oscillator $\hat{H}$ in \eqref{HO} associated with eigenvalue $%
m\in \mathbb{Z}_{+}$ and given by  
\begin{equation}\label{eq5.3}
\phi_m(\xi) :=\left( \sqrt{\pi }2^{m}m!\right) ^{-%
\frac{1}{2}}e^{-\frac{1}{2}\xi ^{2}}H_{m}(\xi ),\qquad\xi \in \mathbb{R}, 
\end{equation}%
where $H_{m}$ is the Hermite polynomial (\cite{Mag}, p.249), as 
\begin{equation}\label{eq5.4}
\left\vert \left( x,y\right) ,\tau ,m\right\rangle :=T_{\tau}(\sigma
_{0}(x,y))[\phi_m] .  
\end{equation}%
Setting $z\equiv x+iy$ and $\tau =2,$ the wavefunction  of the state \eqref{eq5.4}  is given by 
\begin{equation}
\langle \xi \left\vert z,m\right\rangle =\left( -1\right) ^{m}\left( \sqrt{%
\pi }2^{m}m!\right) ^{-\frac{1}{2}}e^{-\frac{1}{2}\overline{z}^{2}+\sqrt{2}%
\xi \overline{z}-\frac{1}{2}\left\vert z\right\vert ^{2}-\frac{1}{2}\xi
^{2}}H_{m}\left( \xi -\frac{z+\overline{z}}{\sqrt{2}}\right) .  
\end{equation}%
These coherent states are completely justified by the square integrability  property  \eqref{eq5.2} of \ $T_{\tau}$ modulo the subgroup $\mathbb{R}$ and the section $\sigma _{0},$ which leads to the resolution of the identity
operator 
\begin{equation}\label{eq5.5}
\mathbf{1}_{L^{2}(\mathbb{R})}=\frac{1}{\pi }\int\limits_{\mathbb{C}%
}d\nu (z)\left\vert z,m\right\rangle \left\langle z,m\right\vert .  
\end{equation}

\noindent As for the canonical CS $\left\vert z\right\rangle \equiv
\left\vert z,0\right\rangle $, Eq. \eqref{eq5.5} allows to implement
a CS quantization of the set of parameters $z$ $\in \mathbb{C}$ by
associating to a function $\mathbb{C\ni }z\mapsto F\left( z,\overline{z}%
\right) \in \mathbb{R}$ the operator on $L^{2}\left( \mathbb{R}\right) :$%
\begin{equation}\label{eq5.6}
F\longmapsto P_{F}:=\frac{1}{\pi }\int\limits_{\mathbb{C}}d\nu (z) \left\vert
z,m\right\rangle \left\langle z,m\right\vert F\left( z,\overline{z}\right). 
\end{equation}%
We can prove (Appendix A) that $P_{F}$ admits the following discrete
resolution with respect to the  orthonormal basis $|j\rangle\equiv \phi_j$ given by  \eqref{eq5.3} as 
\begin{equation}\label{eq5.7}
P_{F}=\sum_{j,k=0}^{+\infty}\mathfrak{a}_{j,k}^{F}\left\vert
j\right\rangle \left\langle k\right\vert  
\end{equation}%
with the matrix elements%
\begin{equation}\label{eq5.9}
\mathfrak{a}_{j,k}^{F}=c_{j,k}^{\left( m\right) }\int\limits_{%
\mathbb{C}}e^{-|z|^{2}}|z|^{|k-m|+|j-m|}e^{i(j-k)\arg z}L_{m\wedge
j}^{(|j-m|)}(|z|^{2})L_{m\wedge k}^{(|k-m|)}(|z|^{2})F (z)d\nu (z),
\end{equation}%
where%
\begin{equation}
c_{j,k}^{\left( m\right) }:=\frac{1}{\pi m!\sqrt{j!k!}} (-1)^{m\wedge j+m\wedge k}(m\wedge
k)!(m\wedge j)! .
\end{equation}%
In particular, when $F$ is a radial function, i.e., $F(z)=\phi (r),$ $%
r=\left\vert z\right\vert $, then, we may use polar coordinates to rewrite \eqref{eq5.9} as

\begin{equation}\label{eq5.11}
\mathfrak{a}_{j,k}^{F}=c_{j,k}^{\left( m\right)
}\int\limits_{0}^{\infty }e^{-r^{2}}r^{|k-m|+|j-m|}L_{m\wedge
j}^{(|j-m|)}(r^{2})L_{m\wedge k}^{(|k-m|)}(r^{2})\phi \left( r\right)
rdr\int\limits_{0}^{2\pi }e^{i(j-k)\theta }d\theta  .
\end{equation}%
From \eqref{eq5.11}, only the coefficient corresponding to $j=k$ is
nonzero and this amounts to the  eigenvalues of $P_{F}$ with the form 
\begin{equation}\label{eq5.12}
\lambda _{k}^{F}:=\mathfrak{a}_{k,k}^{\digamma }=\frac{(m\wedge k)!}{(m\vee
k)!}\int\limits_{0}^{\infty }e^{-\rho }\rho ^{|k-m|}\left( L_{m\wedge
k}^{(|k-m|)}(\rho )\right) ^{2}\phi (\sqrt{\rho })d\rho  
\end{equation}%
which provides the following reduced form of \eqref{eq5.7}: 
\begin{equation}
P_{F}=\sum_{k=0}^{+\infty }\lambda _{k}^{F}\left\vert k\right\rangle
\left\langle k\right\vert  .
\end{equation}%
Moreover, it's not difficult to see that $P_{F}\left\vert k\right\rangle
=\lambda _{k}^{F}\left\vert k\right\rangle ,$ meaning that the Hermite
functions \eqref{eq5.3} are eigenfunctions of $P_{F}$. The latter one may
also be unitarly intertwined, via the coherent states transform $\mathcal{W}%
_{m}:L^{2}(\mathbb{R})\longrightarrow \mathcal{A}_{m}(\mathbb{C})$ defined
by \cite{Mou3} : 
\begin{equation}
\mathcal{W}_{m}[\phi ](z)=(-1)^{m}(2^{m}m!\sqrt{\pi })^{-\frac{1}{2}}e^{-%
\frac{1}{2}\overline{z}^{2}}\int\limits_{\mathbb{R}}\phi (\xi )e^{\sqrt{2}%
\xi \overline{z}-\frac{1}{2}\xi ^{2}}H_{m}\left( \xi -\frac{z+\overline{z}}{%
\sqrt{2}}\right) d\xi,  
\end{equation}%
to act as $\widetilde{P}_{F}=\mathcal{W}_{m}\circ P_{F}\circ \mathcal{W%
}_{m}^{-1}$\ on the $m$-true-polyanalytic space $\mathcal{A}_{m}(\mathbb{C}).$ If $F\equiv \rchi _{D_{R}} $ the indicator
function of the disk $D_{R}$,  $\widetilde{P}_{F}$ turns out to be the concentration operator. That is, the  restriction to the disk $%
D_{R}$ of the integral operator $\tilde{\mathfrak{K}}_{m}$  on $L^{2}(\mathbb{C},d\lambda )$ with the kernel $\widetilde{K}_{m}(z,w)$ in \eqref{eq3.5}, which is the reproducing kernel of  $\mathcal{A}_{m}(\mathbb{C})$.

\section{Mean value and variance of $\sharp _{D_{R}}$}

We here will be dealing with two functions: $\left( i\right) $ the indicator
function $F_R\equiv \rchi _{D_{R}}$ of the disk $D_{R}$ and $\left( ii\right) $
the function 
\begin{equation}\label{eq6.1}
G_R\left( z,\overline{z}\right) \equiv Area\left( D_{R}^{c}\cap D_{R}\left(
z\right) \right) =%
\begin{cases}
\pi R^{2} & ,\left\vert z\right\vert >2R \\ 
\pi R^{2}-2R^{2}\arccos (\frac{1}{2R}\left\vert z\right\vert )+\frac{1}{2}%
\left\vert z\right\vert \sqrt{4R^{2}-\left\vert z\right\vert ^{2}} & 
,\left\vert z\right\vert \leq 2R%
\end{cases}
\end{equation}%
Each of these functions will be treated as quantizable classical observables on the
phase space $\mathbb{C}=\mathbb{R}^{2}$ in order to derive some spectral
properties for their \textit{\ }quantum counterpart\textit{\ }observables which are (essentially) self-adjoint operators on $L^{2}\left( 
\mathbb{R}\right) $, resulting by applying  the coherent states quantization
mapping \eqref{eq5.6}.

\subsection{The number mean of $\sharp _{D_{R}}$}
For $F_R$ $\equiv \rchi_{D_{R}},$ the eigenvalues \eqref{eq5.12} take
the form (see Appendix B):%
\begin{equation}\label{egv1}
\beta _{k}^{(m,R)}=\sum\limits_{j=0}^{2\left( m\wedge k\right) }\mathfrak{a}%
_{j}^{\left( m,k\right) }\gamma \left( \left\vert k-m\right\vert
+j+1,R^{2}\right)  
\end{equation}%
where 
\begin{equation}
\mathfrak{a}_{j}^{\left( m,k\right) }=m!k!\left( -1\right)
^{j}\sum\limits_{\ell =0}^{j}\frac{1}{\ell !\left( j-\ell \right) !\left(
m\wedge k-j+\ell \right) !\left( m\wedge k-\ell \right) !\left( \left\vert
k-m\right\vert +j-\ell \right) !\left( \left\vert k-m\right\vert +\ell
\right) !}  
\end{equation}%
and 
\begin{equation}
\gamma \left( \alpha ,\tau \right) =\int\limits_{0}^{\tau }t^{\alpha
-1}e^{-t}dt,\text{ \ \ }\func{Re}\alpha >0 
\end{equation}%
denotes the incomplete gamma function (\cite{Mag}, p.337). \smallskip\\

For the DPP $\mu _{K_{m},d\nu }$ the number of points $\sharp _{D_{R}},$ that
fall in the disk $D_{R}$  has the same
distribution as a sum of independent Bernoulli random variables $(\beta
_{k}^{(m,R)})_{k\geq 0}$. \ Straightforward calculations (see Appendix C) leads
to the mean value

\begin{equation}
\mathbb{E}\left( \sharp _{D_{R}}\right) =\sum\limits_{k=0}^{+\infty }\beta
_{k}^{(m,R)}=R^{2}  
\end{equation}%
as expected.\medskip\\
\textbf{Remark 6.1.} Note that for $m=0,$ the eigenvalues \eqref{egv1} reduce to 
\begin{equation}
\beta _{k}^{(0,R)}=\frac{1}{k!}\gamma \left( k+1,R^{2}\right)  
\end{equation}%
from which we recover, up to a scaling $R\rightarrow \frac{1}{\sqrt{2}}R,$ the
Daubechies's result (\cite{Daubechies}, p.610).\\

\subsection{The variance of $\sharp _{D_{R}}$}
On one hand, we may apply the above quantization scheme, via the CS $|z,m\rangle$, to the  function $G_R$ taken as a  radial weight function  to construct the  discrete  spectral resolution 
\begin{equation}\label{eq6.7}
P_G=\sum_{k=0}^{+\infty} \lambda_k^{(m,R)} |k\rangle\langle k|
\end{equation}
for the corresponding  operator $P_G$. Indeed, direct calculations (see Appendix D)  gives us the eigenvalues 
\begin{equation*}
\lambda _{k}^{(m,R)}=\left( \pi R\right) ^{2}-\frac{m!\pi
^{3/2}(2R)^{2(k-m)+4}\Gamma ((k-m)+\frac{3}{2})}{4\Gamma (2(k-m)+1)\Gamma
((k-m)+2)k!(k-m+2)(k-m+1)}
\end{equation*}%
\begin{equation}\label{eq6.8}
\times \sum_{s=0}^{2m} A_s^{(m,k)} ~~_{3}F_{3}\left( 
\begin{matrix}
2(k-m)+s+1,(k-m)+\frac{3}{2},(k-m)+1 \\ 
2(k-m)+1,(k-m)+2,(k-m)+3%
\end{matrix}%
;-4R^{2}\right) 
\end{equation}
with 
\begin{equation*}
A_s^{(m,k)}:=\frac{(-1)^{s}(2(k-m)+s)!}{s!}\sum\limits_{r=0}^{s}\left( 
\begin{array}{c}
s \\ 
r%
\end{array}%
\right) \left( 
\begin{array}{c}
k \\ 
m-s+r%
\end{array}%
\right) \left( 
\begin{array}{c}
k \\ 
m-r%
\end{array}%
\right),\qquad k\geq m.
\end{equation*}

 Next,  the variance of $\sharp _{D_{R}}$ counting the number of points
inside the disk $D_{R}$\ can be expressed as a \textquotedblleft \textit{%
dequantization\textquotedblright } of $P_R$ with respect to the set of CS $\left\vert w ,m\right\rangle $ as%
\begin{equation}\label{eq6.82}
\mathbb{V}ar(\sharp _{D_{R}})=\langle w ,m\left\vert P_{R}\right\vert
w ,m\rangle \text{ }\mid _{w =0}  
\end{equation}%
that is the Berezin transform of $G_R$ evaluated at the point $w=0$. So inserting \eqref{eq6.7} into \eqref{eq6.82}, we 
\begin{equation}
\mathbb{V}ar(\sharp _{D_{R}})=\langle 0,m\left\vert \sum_{k=0}^{+\infty
}\lambda _{k}^{\left( m,R\right) }\left\vert k\right\rangle \left\langle
k\right\vert \right\vert 0,m\rangle =\sum_{k=0}^{+\infty }\lambda
_{k}^{\left( m,R\right) }\left( R\right) \langle m\left\vert k\right\rangle
\left\langle k\right\vert m\rangle =\lambda _{m}^{ (m,R) } 
\end{equation}%
Therefore, putting $k=m$ in \eqref{eq6.8}, we obtain%
\begin{equation}\label{eq6.12}
\mathbb{V}ar(\sharp _{D_{R}})=R^{2}\left[ 1-R^{2}%
\sum_{s=0}^{m}(-1)^{s}\left( 
\begin{array}{c}
m \\ 
s%
\end{array}%
\right) {}_{3}F_{2}\left( 
\begin{matrix}
-m,-s,-s \\ 
1,m-s+1%
\end{matrix}%
;-1\right) ~~_{2}F_{2}\left( 
\begin{matrix}
s+1,\frac{3}{2} \\ 
3,2%
\end{matrix}%
;-4R^{2}\right) \right]  .
\end{equation}%
For the first Landau level $m=0$, Eq. \eqref{eq6.12} reduces to 
\begin{equation}\label{eq6.13}
\mathbb{V}ar(\sharp _{D_{R}})=R^{2}\left[ 1-R^{2}%
 ~_{2}F_{2}\left( 
\begin{matrix}
1,\frac{3}{2} \\ 
3,2%
\end{matrix}%
;-4R^{2}\right) \right] .
\end{equation}
By applying the identity (\cite{grad}, p.589):
\begin{equation}
~_{2}F_{2}\left( 
\begin{matrix}
1,\frac{3}{2} \\ 
3,2%
\end{matrix}%
;-4R^{2}\right)=R^{-2} \left[1- e^{-2R^2}\left( I_0(-2R^2)-I_1(-2R^2)    \right)\right]
\end{equation}
in terms of modified Bessel functions  $I_0$ and $I_1$, Eq. \eqref{eq6.13}  also reads 
\begin{equation}
\mathbb{V}ar(\sharp _{D_{R} })= R^2e^{-2R^2}\left( I_0(2R^2)+I_1(2R^2)\right)
\end{equation}
which corresponds to the (infinite) Ginibre process in agreement  with  the result of Osada and Shirai (\cite{Osa-shi}, p.2). 
\section*{Appendix A}
In order to prove \eqref{eq5.7}, we need to show that the operator $P_F$ defined by \eqref{eq5.6} satisfies 
\begin{equation}
\langle P_F[\phi_j],\phi_k\rangle =\mathfrak{a}_{j,k}^F\tag{A1}
\end{equation}
where $\mathfrak{a}_{j,k}^F$ are the coefficients given by \eqref{eq5.9}. Here $\{\phi_j\}$ is the basis vector of Hermite functions defined in \eqref{eq5.3}. For this, we may use the number states expansion of the CS $|z,m\rangle$   (\cite{Mou3}, p.4) :

\begin{equation}
|z,m\rangle=\left(e^{|z|^{2}}\right)^{-\frac{1}{2}} \sum_{j \geq 0} \frac{(-1)^{m}}{\sqrt{\pi m ! j !}} m !|z|^{(j-m)} e^{i(m-j)  \arg z} L_{m}^{j-m}\left(|z|^{2}\right) \phi_{j}\tag{A2}
\end{equation}
to write the above coefficients as 
\begin{eqnarray*}
\hspace*{-3cm}\mathfrak{a}_{j, k}^{\digamma}&=&\left\langle \frac{1}{\pi}\int_{\mathbb{C}}|z, m\rangle\langle z,m| F(z, \bar{z}) d\nu(z)[\phi_j],\phi_k\right\rangle_{L^{2}(\mathbb{R})}\cr
&=&  \frac{1}{\pi} \int_{\mathbb{C}}  \left\langle \phi_j|z, m\right\rangle\left\langle z,m|\phi_k\right\rangle F(z, \bar{z})d\nu(z)
\end{eqnarray*}
\begin{equation}
= \frac{1}{\pi^2}\frac{m !}{\sqrt{j ! k !}} \int_{\mathbb{C}}|z|^{(j+k-2 m)} e^{-|z|^{2}} e^{i(j-k) \arg z} L_{m}^{(j-m)}\left(|z|^{2}\right) L_{m}^{(k-m)}\left(|z|^{2}\right) F(z, \bar{z}) d \nu(z) \tag{A3}
\end{equation}
which completes the proof of Appendix A. \qquad $\square$
\section*{Appendix B}
For $F=\rchi_{D_R}$ the indicator function of the disk $D_R$, the eigenvalues  \eqref{eq5.12} take the form
\begin{equation}\label{B1}
\beta _{k}^{(m,R)}=\frac{(m\wedge k)!}{(m\vee k)!}\int%
\limits_{0}^{R^{2}}e^{-\rho }\rho ^{|k-m|}\left( L_{m\wedge
k}^{(|k-m|)}(\rho )\right) ^{2}d\rho.  \tag{B1}
\end{equation}%
Making use of the Feldheim's formula (\cite{Feld}, Eq.1.13)
:
\begin{equation}
L_{q}^{\left( \alpha \right) }\left( x\right) L_{p}^{\left( \alpha \right)
}\left( x\right) =\sum\limits_{j=0}^{q+p}\left( -1\right) ^{j}A_{j}^{\left(
q,p,\alpha \right) }\frac{x^{j}}{j!}  \tag{B2}
\end{equation}%
where%
\begin{equation}
A_{j}^{\left( q,p,\alpha \right) }=\sum\limits_{\ell =0}^{j}\left( 
\begin{array}{c}
j \\ 
\ell%
\end{array}%
\right) \left( 
\begin{array}{c}
q+\alpha \\ 
q-j+\ell%
\end{array}%
\right) \left( 
\begin{array}{c}
p+\alpha \\ 
p-\ell%
\end{array}%
\right)  \tag{B3}
\end{equation}%
for parameters $p=q=m\wedge k,$ $\alpha =\left\vert k-m\right\vert $ and $%
x=\rho $, then we may write 
\begin{equation}\label{B4}
\frac{(m\wedge k)!}{(m\vee k)!}\left( L_{m\wedge k}^{(|k-m|)}(\rho )\right)
^{2}=\sum\limits_{j=0}^{2\left( m\wedge k\right) }\mathfrak{a}_{j}^{\left(
m,k\right) }\rho ^{j}.  \tag{B4}
\end{equation}%
By inserting  \eqref{B4} into \eqref{B1}, we obtain 
\begin{equation}
\beta _{k}^{(m,R)}=\sum\limits_{j=0}^{2\left( m\wedge k\right) }\mathfrak{a}%
_{j}^{\left( m,k\right) }\int\limits_{0}^{R^{2}}e^{-\rho }\rho
^{|k-m|+j}d\rho =\sum\limits_{j=0}^{2\left( m\wedge k\right) }\mathfrak{a}%
_{j}^{\left( m,k\right) }\gamma \left(| k-m|+j-1,R^{2}\right)  \tag{B5}
\end{equation}%
as announced in \eqref{egv1}.\qquad $\square$
\section*{Appendix C}
The random variable $\sharp_{D_R}$ has the same distribution as a sum of independent Bernoulli random variables of parameters $0<\beta_{k}^{(m, R)}<1$, $k\geq 0$. Therefore, the expectation of $\sharp_{D_R}$ is given by the sum 
\begin{equation}\label{C1}
\mathbb{E}\left( \sharp _{D_{R}}\right)=\sum_{k =0}^{+\infty} \beta_{k}^{(m, R)}=\sum_{k \geqslant 0} \frac{(m \wedge k) !}{(m \vee k) !} \int_{0}^{R^{2}} e^{-\rho} \rho^{|k-m|}\left(L_{m \wedge k}^{|k-m|}(\rho)\right)^{2} d \rho.\tag{C1}
\end{equation}
The integer $m$ being fixed, we may write the series
\begin{equation}
\mathfrak{S}_m(\rho)=\sum_{k =0}^{+\infty} \frac{(m \wedge k) !}{(m \vee k) !} \rho^{|k-m|}\left(L_{m \wedge k}^{|k-m|}(\rho)\right)^{2} \tag{C2}
\end{equation}
 as $S_{<\infty}(m,\rho)+S_{\infty}(m,\rho)$, where 
\begin{equation}
S_{<\infty}(m,\rho)=\sum_{k =0}^{m-1} \frac{k !}{m  !} \rho^{m-k}\left(L_{ k}^{m-k}(\rho)\right)^{2}-\sum_{k =0}^{m-1} \frac{m!}{k!} \rho^{k-m}\left(L_{m}^{k-m}(\rho)\right)^{2}\tag{C3}
\end{equation}
and 
\begin{equation}\label{C4}
S_{\infty}(m,\rho)=\sum_{k =0}^{+\infty} \frac{m!}{k!} \rho^{k-m}\left(L_{m}^{k-m}(\rho)\right)^{2}. \tag{C4}
\end{equation}
By making use of the identity (\cite{zego}, p. 98):
\begin{equation}
L_{m}^{(k-m)}\left(\rho\right)=\left(-\rho\right)^{m-k} \frac{k !}{m !} L_{k}^{(m-k)}\left(\rho\right),\tag{C5}
\end{equation}
one can check  that $S_{<\infty}(m,\rho)=0$. For the infinite sum in \eqref{C4}, we may apply the following  formula of Bateman (\cite{Bat}, p.457):
\begin{equation}
\sum_{l=0}^{+\infty} \frac{n !}{l !}\left(\sqrt{x y} e^{i \varphi}\right)^{l-n} L_{n}^{(l-n)}\left(x\right) L_{n}^{(l-n)}\left(y\right)  =\exp \left(\sqrt{x y} e^{i \varphi}\right) L_{n}^{(0)}\left(x+y-2 \sqrt{x y}\cos \varphi\right)
\tag{C6}
\end{equation}
for $n=m,\, x=y=\rho$ and $\varphi=0$. This leads to $S_{\infty}(m,p)=e^{\rho}$. Summarizing the above calculations, Eq. \eqref{C1} reads 
\begin{equation}
\mathbb{E}\left( \sharp _{D_{R}}\right)=\int_{0}^{R^{2}} e^{-\rho} \mathfrak{S}_m(\rho) d \rho=\int_{0}^{R^{2}} d \rho=R^{2}.\qquad\tag{C7}
\end{equation}
This completes the proof of Appendix C.
\section*{Appendix D}
By choosing the radial weight function \eqref{eq6.1}, the eigenvalues  in Eq.\eqref{eq5.12} can be decomposed into three integrals as 
\begin{equation}\label{D1}
\lambda _{k}^{(m,R)}=\sigma_1-\sigma_2+\sigma_3\tag{D1}
\end{equation}
where 
\begin{equation}\label{D2}
\sigma_1:=\pi R^2\frac{m!}{k!}\int\limits_{0}^{\infty }e^{-\rho }\rho ^{k-m}\left( L_{m}^{(k-m)}(\rho )\right) ^{2}d\rho,\tag{D2} 
\end{equation}
\begin{equation}\label{D3}
\sigma_2:=2R^2 \int_{0}^{2R} \rho^{2 (k-m)+1} \arccos \left(\frac{\rho}{2R}\right) e^{-\rho^{2}}\left(L_{m\wedge k}^{(k-m)}\left(\rho^{2}\right)\right)^{2} d \rho,\tag{D3}
\end{equation}
and 
\begin{equation}\label{D4}
\sigma_{3}:=\frac{1}{2} \int_{0}^{2 R} \sqrt{4 R^{2}-\rho^{2}} e^{-\rho^{2}} \rho^{2(k-m)+2}\left(L_{m}^{(k-m)}\left(\rho^{2}\right)\right)^{2} d \rho.\tag{D4}
\end{equation}
To calculate $\sigma_1$, we make use of  the orthogonality relation of Laguerre polynomials  (\cite{Grad}, p.809), the integral \eqref{D2} reduces to $\sigma_1=\pi R^2.$\smallskip\\

For $\sigma_2$, we set $a=2R$ and  $\alpha=k-m$, then
\begin{equation}\label{D5}
\sigma_{2}=\frac{a^{2}}{2} \int_{0}^{a} \rho^{2 \alpha+1} \arccos \left(\frac{\rho}{a}\right) e^{-\rho^{2}}\left(L_{m}^{(\alpha)}\left(\rho^{2}\right)\right)^{2} d \rho .\tag{D5}
\end{equation} 
Next, we apply the Feldheim's formula (\cite{Feld}, Eq.1.14):
\begin{equation}\label{feldid}
L_{m}^{(\alpha)}(x) L_{m}^{(\alpha)}(x)=\sum_{s=0}^{2m} C_{s}(m,\alpha) L_{s}^{(2\alpha)}(x)\tag{D6}
\end{equation}
with
\begin{equation}\label{D7}
C_{s}(m,\alpha)=(-1)^{s} \sum_{r=0}^{s}\left(\begin{array}{l}
s \\
r
\end{array}\right)\left(\begin{array}{c}
m+\alpha \\
m-s+r
\end{array}\right)\left(\begin{array}{c}
m \\
m-r
\end{array}\right). \tag{D7}
\end{equation}
This gives
\begin{equation}\label{D8}
\sigma_{2}=\frac{a^{2}}{2} \sum_{s=0}^{2 m} C_{s}(m,\alpha) \int_{0}^{a} \rho^{2 \alpha+1} \arccos \left(\frac{\rho}{a}\right) e^{-\rho^{2}} L_{s}^{(2 \alpha)}\left(\rho^{2}\right) d \rho. \tag{D8}
\end{equation}
By using the expression of the Laguerre polynomials $ L_{s}^{(2 \alpha)}\left(u^{2}\right)=\sum_{l=0}^{s}\frac{(-1)^{l}}{l!}\left(\begin{array}{c}
s+2 \alpha \\
s-l
\end{array}\right) u^{2l}$ and setting $\rho=ta$, Eq. \eqref{D8} becomes  
\begin{equation}\label{D9}
\sigma_{2}=\frac{a^{2(\alpha+2)}}{2} \sum_{s=0}^{2 m} C_{s}(m,\alpha)  \sum_{l=0}^{s}\frac{(-1)^{l}a^{2l}}{l!}\left(\begin{array}{c}
s+2 \alpha \\
s-l
\end{array}\right) \sigma_2' \tag{D9}
\end{equation}
where 
\begin{equation}
\sigma_2':=\int_{0}^{1} t^{2 (\alpha+l)+1} \arccos \left(t\right) e^{-a^{2}t^2} d t. \tag{D10}
\end{equation}
In order to calculate $\sigma_2'$, we set $u=\arccos (t)$ , $d v=t^{2 (\alpha+l)+1} e^{-a^{2} t^{2}}$ and we can easily check that 
\begin{equation}
v  =\frac{a^{-2 (\alpha+l)-2}}{2} \gamma\left(\alpha+l+1, a^{2} t^{2}\right)  \tag{D11}
\end{equation}
where $\gamma$ is the incomplete Gamma function. Next, we  apply the integration by parts to get that
\begin{equation}\label{D12}
\sigma_{2}' =\frac{a^{-2(\alpha+l)-2}}{2} \int_{0}^{1} \frac{\gamma\left(\alpha+l+1, a^{2} t^{2}\right)}{\sqrt{1-t^{2}}} d t. \tag{D12}
\end{equation}
Performing again  the change of variables $t^{2}=x$, we may rewrite \eqref{D12} as
\begin{equation}
\sigma_{2}'=\frac{a^{-2(\alpha+l)-2}}{8} \int_{0}^{1} x^{-\frac{1}{2}}(1-x)^{-\frac{1}{2}} \gamma\left(\alpha+l+1, a^{2} x\right) d x \tag{D13}
\end{equation}
By applying the following formula  (\cite{grad}, p.143) :
\begin{equation}
\int_{0}^{\zeta} x^{n-1}\left(\zeta^{r}-x^{r}\right)^{\beta-1} \gamma(\nu, c x) d x=\frac{\zeta^{n-r \beta-r+\nu} c^{\nu}}{r} \Gamma(\beta) \sum_{s=0}^{\infty} \frac{(-\zeta c)^{s}}{s !(\nu+s)} \frac{\Gamma((n+\nu+s) / r)}{\Gamma((n+\nu+s) / r+\beta)}  \tag{D14}
\end{equation}
for the parameters $n=1 / 2$, $\zeta=1$, $\beta=1/2$, $\nu=\alpha+l+1$, $r=1$, $c=a^{2}$, to get that 
\begin{equation}
\sigma_{2}'=\frac{\Gamma(1 / 2)}{8} \sum_{s=0}^{\infty} \frac{\left(-a^{2}\right)^{s}}{s !(\alpha+l+1+s)} \frac{\Gamma(s+\alpha+l+3 / 2)}{\Gamma(s+\alpha+l+2)} \tag{D15}
\end{equation}
which can be expressed as
\begin{equation}\label{D16}
\sigma_{2}'=\frac{\sqrt{\pi}}{8} \frac{\Gamma(\alpha+l+3 / 2)}{(\alpha+l+1)^{2} \Gamma(\alpha+l+1)}{ }_{2} F_{2}\left(\begin{array}{c}
\alpha+l+1, \alpha+l+3 / 2 \\
\alpha+l+2, \alpha+l+2 ;-a^{2}
\end{array}\right). \tag{D16}
\end{equation}
By inserting \eqref{D16} into \eqref{D9}, we obtain
\begin{equation}
\sigma_2=\frac{a^{2}\sqrt{\pi}}{2} \sum_{s=0}^{2 m} C_{s}(m,\alpha) K_{l,s}(\alpha) \tag{D17}
\end{equation}
 where
\begin{equation*}
 K_{l,s}(\alpha)= a^{2 \alpha+2} \frac{\Gamma(\alpha+1) \Gamma\left(\alpha+\frac{3}{2}\right)}{\Gamma(\alpha+2) \Gamma(\alpha+2)} \frac{(s+2 \alpha) !}{s ! \Gamma(2 \alpha+1)}
 \end{equation*}
 \begin{equation}\label{D18}
 \times\sum_{l=0}^{s}\left(\begin{array}{l}
s \\
l
\end{array}\right) \frac{\left(-a^{2}\right)^{l}}{(2 \alpha+1)_{l}} \frac{(\alpha+1)_{l}\left(\alpha+\frac{3}{2}\right)_{l}}{(\alpha+2)_{l}(\alpha+2)_{l}}{ }_{2} F_{2}\left(\begin{array}{c}
\alpha+l+1, \alpha+l+3 / 2 \\
\alpha+l+2, \alpha+l+2 ;-a^{2}
\end{array}\right).\tag{D18}
 \end{equation}
By appealing  the identity  (\cite{grad}, p.391) :
\begin{equation}
\sum_{k=0}^{n}\left(\begin{array}{l}
n \\
k
\end{array}\right) \frac{x^{k}}{(\beta)_{k}} \frac{\prod\left(a_{p}\right)_{k}}{\prod\left(b_{q}\right)_{k}}{ }_{p} F_{q}\left(\begin{array}{c}
\left(a_{p}\right)+k ; x \\
\left(b_{q}\right)+k
\end{array}\right)={ }_{p+1} F_{q+1}\left(\begin{array}{c}
\left(a_{p}\right), \beta+n ; x \\
\left(b_{q}\right), \beta
\end{array}\right),\tag{D19}
\end{equation}
 the sum in \eqref{D18} takes the form
\begin{equation}
{ }_{3} F_{3}\left(\begin{array}{c}
\alpha+1, \alpha+\frac{3}{2}, 2 \alpha+1+s ;-a^{2} \\
\alpha+2, \alpha+2,2 \alpha+1
\end{array}\right). \tag{D20}
\end{equation}
Therefore,
\begin{equation}
 K_{l,s}(\alpha)=a^{2 \alpha+2} \frac{\Gamma(\alpha+1) \Gamma\left(\alpha+\frac{3}{2}\right)}{\Gamma(\alpha+2) \Gamma(\alpha+2)} \frac{(s+2 \alpha) !}{s ! \Gamma(2 \alpha+1)}{ }_{3} F_{3}\left(\begin{array}{c}
\alpha+1, \alpha+\frac{3}{2}, 2 \alpha+1+s ;-a^{2} \\
\alpha+2, \alpha+2,2 \alpha+1
\end{array}\right),\tag{D21}
\end{equation} 
and our integral $\sigma_2$ can be expressed as
\begin{equation*}
\sigma_{2}=\frac{(2R)^{2} \sqrt{\pi}}{8} \frac{(2R)^{2 (k-m)+2} \Gamma(k-m+1) \Gamma\left(k-m+\frac{3}{2}\right)}{\Gamma(2 (k-m)+1) \Gamma(k-m+2) \Gamma(k-m+2)} 
\end{equation*} 
\begin{equation}
\times\sum_{s=0}^{2 m} C_{s}(m, k-m) \frac{(s+2(k-m)) !}{s !}{ }_{3} F_{3}\left(\begin{array}{c}
k-m+1, k-m+\frac{3}{2}, 2(k-m)+1+s ;-(2R)^{2} \\
k-m+2, k-m+2,2 (k-m)+1
\end{array}\right).\tag{D22}
\end{equation}
For the last integral, we   again use Feldheim's formula  \eqref{feldid}, then Eq.\eqref{D4} becomes
\begin{equation}\label{D23}
\sigma_{3}=\frac{1}{2} \sum_{s=0}^{2 m} C_{s}(m,k-m) \int_{0}^{2R} \sqrt{4R^{2}-\rho^{2}} e^{-\rho^{2}} \rho^{2(k-m)+2} L_{s}^{(2(k-m))}\left(\rho^{2}\right) d \rho\tag{D23}
\end{equation}
where $C_{s}(m,k-m)$ is given by \eqref{D7}. Setting  $\alpha=k-m$, $\rho^{2}=(2R)^{2} t,$ we may rewrite \eqref{D23} as 
\begin{equation}
\sigma_3 =\frac{(2R)^{2\alpha+4}}{4} \sum_{s=0}^{2 m} C_{s}(m,\alpha) \int_{0}^{1} t^{\alpha+\frac{1}{2}} e^{-(2R)^{2} t}(1-t)^{\frac{1}{2}} L_{s}^{(2\alpha)}\left((2R)^{2} t\right) d t .\tag{D24}
\end{equation}
Now, we use the  formula (\cite{Grad}, p.810):
\begin{equation}
\int_{0}^{1}(1-x)^{\mu-1} x^{\lambda-1} e^{-\beta x} L_{s}^{(\gamma)}(\beta x) d x=\frac{\Gamma(\gamma+s+1)}{s ! \Gamma(\gamma+1)}\frac{\Gamma(\lambda)\Gamma(\mu)}{\Gamma(\lambda+\mu)} {}_2F_{2}\left(\begin{array}{c}
\gamma+s+1, \lambda \\
\gamma+1, \lambda+\mu  ;-\beta
\end{array}\right)   \tag{D25}
\end{equation}
for the parameters $\mu =\frac{3}{2}$, $\lambda =\alpha+\frac{3}{2}$, $\beta=(2R)^{2}$, $\gamma =2\alpha$,  to get that
\begin{equation}\label{D26}
\sigma_{3}=\frac{(2R)^{2\alpha+4}\Gamma(\alpha+\frac{3}{2})\Gamma(\frac{3}{2})}{4\Gamma(2\alpha+1)\Gamma(\alpha+3)}   \sum_{s=0}^{2 m} C_{s}(m,\alpha)  \frac{\Gamma(2\alpha+s+1)}{s !}{}_2F_{2}\left(\begin{array}{c}
2\alpha+s+1,\alpha+\frac{3}{2} \\
 2\alpha+1,\alpha+3  ;-(2R)^{2}
\end{array}\right).\tag{D26}
\end{equation}
Next, to compute the difference $\sigma_3-\sigma_2$ we may apply the following identity 
\begin{equation*}
\frac{1}{(d+2)}{ }_{p} F_{q}\left(\begin{array}{c}
a_{1}, a_{2}, \ldots, a_{p-1}, a_{p} \\
b_{1}, b_{2}, \ldots, b_{q-1}, d+3
\end{array} ; z\right)-\frac{1}{(d+1)} {}_{p+1}F_{q+1}\left(\begin{array}{c}
a_{1}, a_{2}, \ldots, a_{p-1}, a_{p}, d+1 \\
b_{1}, b_{2}, \ldots, b_{q-1}, d+2, d+2
\end{array} ; z\right) 
\end{equation*}
\begin{equation}\label{D27}
 =-\frac{1}{(d+2)(d+1)}{ }_{p+1} F_{q+1}\left(\begin{array}{c}
a_{1}, a_{2}, \ldots, a_{p-1}, a_{p}, d+1 \\
b_{1}, b_{2}, \ldots, b_{q-1}, d+2, d+3
\end{array} ; z\right),\qquad p,q\in\mathbb{N},\; z \in \mathbb{C}.\tag{D27}
\end{equation}
which is easy to check  from the definition of the hypergeometric series ${}_pF_q$.
Finally, we obtain, after summarizing all the calculations in \eqref{D1}, the expression of the eigenvalues given by \eqref{eq6.8}.\quad $\square$


\begin{thebibliography}{99}
\bibitem{Hough}  J. B. Hough, M. Krishnapur, Y. Peres, B. Virág, Zeros of Gaussian analytic functions and determinantal point processes, University Lecture Series. \textbf{51}. American Mathematical Society, Providence (2009).
\bibitem{Soch} A. Soshnikov, Determinantal Random Point Fields. \textit{Russian Math. Surveys}. \textbf{55} (2000), 923-975.

\bibitem{Kulesza-Taskar} A. Kulesza,  and B. Taskar, Determinantal
point processes for machine learning. \textit{Foundations and Trends in Machine Learning.} \textbf{5} (2012). 123-286.
\bibitem{Leo} G. L. Torrisi and E. Leonardi, Large deviations of the interference in the Ginibre network model.
\textit{Stochastic Systems}, \textbf{4} (2014), 1–33.
\bibitem{Miyoshi-Shirai} N. Miyoshi and T. Shirai,  A Cellular Network Model with Ginibre Configured Base Stations. \textit{Advances in Applied Probability}, \textbf{46} (2014), 832-845.
\bibitem{Macchi} {O. Macchi}, The coincidence approach to stochastic
point processes, \textit{Adv. Appl. Prob.} \textbf{7} (1975), 83-122.
\bibitem{Torquato} S. Torquato, A. Scardicchio, and C. E. Zachary, Point
Processes in Arbitrary Dimension from Fermionic Gases,
\textit{Random Matrix Theory, and Number Theory, J. Stat.
Mech.} (2008). P11019. 
\bibitem{Shirai1} T. Shirai,  Large Deviations for the Fermion Point Process
Associated with the Exponential Kernel. \textit{J. Stat. Phys.} \textbf{123} (2006), 615-629.


\bibitem{Ginibre} J. Ginibre,  Statistical Ensembles of Complex, Quaternion,
and Real Matrices. \textit{J. Math. Phys}. \textbf{6} (1965), 440-449. 

\bibitem{JLM} B. Jancovici, J. L. Lebowitz, G. Manificat, Large charge
fluctuations in classical Coulomb systems.\textit{ J. Statist. Phys. } \textbf{72} (1993), 773-787.

\bibitem{AIM} {N. Askour, A. Intissar, Z. Mouayn}, Espaces de
Bargmann g\'{e}n\'{e}ralis\'{e}s et formules explicites pour leurs noyaux
reproduisants, \textit{C. R. Acad. Sci. Paris S\'{e}rieI. Math.} \textbf{325} (1997), 707-712.

\bibitem{Shirai2} {T. Shirai}, Ginibre-type point processes and
their asymptotic behavior, \textit{J. Math. Soc. Japan.} \textbf{67} (2015), 763-787.


\bibitem{abreu17} L. D. Abreu, J. Pereira, J. L. Romero and S.  Torquato, The Weyl-Heisenberg ensemble: Hyperuniformity and higher Landau levels. \textit{ J. Stat. Mech. Theor. Exp.} \textbf{4} (2017). 



\bibitem{Lenard1} {A. Lenard}, Correlation functions and the
uniqueness of the state in classical mechanics, \textit{Commun. Math. Phys.} \textbf{30} (1973),
35-44.

\bibitem{Lenard2} {A. Lenard}, States of classical statistical
mechanical system of infinitely many particles I,\textit{ Arch. Rationnal. Mech.
Anal.} \textbf{59} (1975), 219-239.
\bibitem{Mag} {W. Magnus, F. Oberhettinger, R. P. Soni},  Formulas
and Theorems for the special Functions of Mathematical Physics, 3rd edn.
\textit{Die Grundlehren der mathematischen Wissenschaften}, vol. \textbf{52}. Springer, New
York (1966).

\bibitem{DL} N. Demni and P. Lazag. The hyperbolic-type point process. \textit{J. Math. Soc. Japan}. \textbf{71} (2019), 1137-1152.


\bibitem{schro} E. Schrodinger, Der stretige Ubergang von der Mikro-zur
Makromechanik, \textit{Naturwissenschaften}. \textbf{14} (1926), 664-666. 

\bibitem{RS} M. Reed and B. Simon,  Methods of Modern Mathematical Physics,
Vols. I-IV \symbol{126}Academic, New York, 1978.

\bibitem{GO} J. P. Gazeau, M. A. del Olmo, Pisot \textit{q}-coherent states quantization of the harmonic oscillator,
\textit{Annals of Physics}, \textbf{330} (2013), 220-245.
\bibitem{Ga} J. P. Gazeau, Coherent states in quantum physics, WILEY-VCH
Verlag GMBH \& Co. KGaA Weinheim 2009.

\bibitem{K2} J. R. Klauder,. Quantization without quantization. \textit{Ann. Physics}. \textbf{237} (1995), 147-160. .

\bibitem{K3} J. R. Klauder, \ Beyond conventional quantization, Cambridge
University Press, Cambridge 2000.

\bibitem{Ber} F. A. Berezin, General concept of quantization, \textit{Comm. Math.
Phys.}, \textbf{40} (1975), 153-174.

\bibitem{Fol} G. B. Folland, Harmonic analyse on phase space, in Annals of
Math Studies No.122, Princeton U.P., Princeton NJ (1989).

\bibitem{Taylor} M. E. Taylor, Noncommutative Harmonic Analysis, Vol \textbf{22},
Mathematical Survey and monographs, AMS, Providence, RI (1986). 

\bibitem{BJR} C. Benson, J. Jenkins and G. Ratclif, Bounded $K-$spherical
functions on Heisenberg groups, \textit{J. Funct. Ann}, \textbf{105} (1992), 405-443. 


\bibitem{Duflo} M. Duflo  and C. C. Moore, On the regular representation of a
Nonunimodular locally compact group, \textit{J. Funct. Anal,} \textbf{21}  (1976), 209-243.

\bibitem{BPS} E. Binz, S. Pods and W. Schemp, Heisenberg groups-the fundamental
ingredient to describe information, its transmission and quantization, \textit{J.
Phys. A: Math. Gen.} \textbf{36} (2003), 6401-6421. 


\bibitem{Mou3} Z. Mouayn, Coherent state transforms attached to generalized
Bargmann spaces on the complex plane,\textit{ Math. Nachr.} \textbf{284} (2011), 1948-1954.

\bibitem{Perelomov} A. Perelomov, Coherent States for Arbitrary Lie group, 
\textit{Commun. Math. Phys}. \textbf{26} (1972), 222-236. 







\bibitem{Daubechies} I. Daubechies, Time-frequency localization operators: a
geometric phase space approach, \textit{IEEE transactions on information theory}, \textbf{34} (1998).

\bibitem{grad} A. P. Prudnikov, Yu. A. Brychkov and O. I. Marichev. Integrals and Series-More special functions. Volume \textbf{3}. 1986.

\bibitem{Osa-shi} H. Osada and T. Shirai, Variance of the linear statistics of the Ginibre random point field,
\textit{RIMS Kôkyûroku Bessatsu}, \textbf{B6} (2008), 193-200.

\bibitem{Feld} E. Feldheim, Expansions and integral transforms for products of Laguerre and Hermite polynomials, \textit{Quarterly Journal of Mathematics (Oxford)}, \textbf{11} (1940), 18–29.
\bibitem{zego} G. Szego, Orthogonal Polynomials (American Mathematical Society, Providence, R.I., 1975).
\bibitem{Bat} H. Bateman, Partial Differential Equations of Mathematical Physics, Cambridge, 1932.



\bibitem{Grad}I. Gradshteyn and I. Ryzhik. Table of integrals, series, and products. Academic Press, Amsterdam, Seventh edition, (2007).



\end{thebibliography}
\end{document}